\documentclass[usenatbib]{mn2e}

\usepackage{url, amssymb, textcomp, verbatim, graphicx, times}
\usepackage[usenames, dvipsnames]{color}

\newcommand{\adeg}[1]{{#1}$^{\circ}$}
\newcommand{\amin}[1]{{#1}$^\prime$}
\newcommand{\asec}[1]{{#1}$^{\prime\prime}$}
\newcommand{\thour}[1]{{#1}$^{\mathrm{h}}$}
\newcommand{\tmin}[1]{{#1}$^{\mathrm{m}}$}
\newcommand{\tsec}[1]{{#1}$^{\mathrm{s}}$}

\newcommand{\mjybeam}[1]{{#1}\,mJy\,beam$^{-1}$}
\newcommand{\mujybeam}[1]{{#1}\,$\mu$Jy\,beam$^{-1}$}

\newcommand{\hms}[3]{\thour{#1}\tmin{#2}\tsec{#3}}
\newcommand{\dms}[3]{\adeg{#1}\amin{#2}\asec{#3}}

\newcommand{\degrees}{\ifmmode^{\circ}\else$^{\circ}$\fi}
\newcommand{\degree}{\ifmmode^{\circ}\else$^{\circ}$\fi}

\newcommand{\fermi}{\emph{Fermi} }

\def\aj{AJ}
\def\apj{ApJ}
\def\apjl{ApJ}
\def\apjs{ApJS}
\def\aap{A\&A}
\def\mnras{MNRAS}
\def\nat{Nature}

\title{Searching For Pulsars Associated With the Fermi GeV Excess}

\author[Bhakta et al.]{D. Bhakta$^{1,2}$, J. Deneva$^{3}$, D. A. Frail$^2$, F. de Gasperin$^4$,
H. T. Intema$^4$, 
\newauthor{P. Jagannathan$^{2,5}$ and K.~P.~Mooley$^{6,7}$}
\\
$^{1}$ Department of Physics, Texas Tech University, Box
   41051, Lubbock, TX 79409-1051, USA\\
  $^{2}$ National Radio Astronomy Observatory, 1003 Lopezville Road, Socorro, NM 87801, USA; Email: dfrail@nrao.edu\\
  $^{3}$ George Mason University, resident at Naval Research Laboratory, Washington, DC 20375, USA\\
$^{4}$Leiden Observatory, Leiden University, Niels Bohrweg 2, NL-2333CA, Leiden, The Netherlands\\
  $^{5}$ Department of Astronomy, University of Cape Town, Private Bag X3, Rondebosch 7701, Republic of South Africa\\
  $^{6}$ Astrophysics, Department of Physics, University of Oxford, Keble Road, Oxford OX1 3RH, UK \\
  $^{7}$ Hintze Research Fellow\\
}

\begin{document}
\maketitle

\begin{abstract}

The Fermi Large Area Telescope has detected an extended region of GeV emission toward the Galactic Center that is currently thought to be powered by dark matter annihilation or a population of young and/or millisecond pulsars. In a test of the pulsar hypothesis, we have carried out an initial search of a 20 deg$^2$ area centered on the peak of the galactic center GeV excess. Candidate pulsars were identified as a compact, steep spectrum continuum radio source on interferometric images and followed with targeted single-dish pulsation searches. We report the discovery of the recycled pulsar PSR\,1751$-$2737 with a spin period of 2.23 ms.  PSR\,1751$-$2737 appears to be an isolated recycled pulsar located within the disk of our Galaxy, and it is not part of the putative bulge population of pulsars that are thought to be responsible for the excess GeV emission. However, our initial success in this small pilot survey suggests that this hybrid method (i.e. wide-field interferometric imaging followed up with single dish pulsation searches) may be an efficient alternative strategy for testing whether a putative bulge population of pulsars is responsible for the GeV excess.

\end{abstract}


\section{Introduction}

Observations by the \textit{Fermi} Large Area Telescope (LAT) have revealed an excess of GeV emission from the center of our Galaxy \citep[See review by][]{vane15}. This diffuse emission can be seen over a 10$^\circ\times 10^\circ$ region toward the galactic center but is strongest within a 2.5$^\circ$ radius of Sgr A* \citep{aaa+16}. There are currently two alternative explanations which have been offered to explain this diffuse excess: (a) it is the long-sought annihilation signature of dark matter particles \citep{vm09,hg11,wei12,dfh+16}, or (b) the integrated high energy emission from a population of several thousand young and/or millisecond pulsars \citep{aba11,bk15}. Recent analysis of the spatial and spectral properties of the gamma-ray excess strongly favor the pulsar hypothesis \citep{aba11,mir13,yz14,ccw15,aaa+16,lls+16,bkw16}.

The essential test of the pulsar hypothesis would be to detect individual bulge pulsars and show that their properties are consistent with the GeV excess signal. It has long been argued from theoretical grounds and multi-wavelength observations that a substantial population of pulsars exists both in close orbit around the black hole Sgr A* and on larger scales of 100's of parsecs around the galactic center \citep{pl04,mpb+05,wlg06,wcc+12}. However, despite extensive searches for radio pulsations, only a handful of normal (i.e. non-recycled) pulsars and one magnetar have been discovered within a few hundred parsecs of the galactic center, but no millisecond pulsars \citep{jkl+06,dcl09,sj13,mgz+13,efk+13}.

The observational challenges in finding the putative bulge pulsars responsible for the gamma-ray excess have recently been summarized by \citet{cdd+16} and \citet{lkkd16}. The large distance and the high brightness temperature of the diffuse synchrotron emission toward the bulge means that any pulsed signals will be weak and hard to detect without deep integrations. More problematic, however, is the large amount of ionized gas at the galactic center, which results in heavy scattering and dispersion broadening \citep{fdcv94,lc98,mk15}. The effects are strongest within the central 150 pc radius of Sgr A*, but dispersive smearing and scattering remain large over most of area of the GeV excess \citep[see Figs. 5 and 6 of][]{cdd+16}. Temporal smearing of pulsed signals make them hard to detect unless observing is carried out at high frequencies ($\nu>5$ GHz) where temporal scattering is sharply reduced ($\tau_{scat}\propto\nu^{-4}$). This approach has been used for targeted radio pulsar searches in the immediate vicinity of SgrA* \citep[e.g.][]{mkfr10,ekk+13,sbb+13} but it is not feasible today to carry out a blind pulsation survey over the entire region of the GeV excess.

An alternative, potentially more efficient, approach is to begin by initially identifying candidates in the image plane \citep{cl97,lc08}, and then later followed up with deep radio pulsations searches at higher frequencies. Pulsar candidates can be recognized in the image plane as compact, steep spectrum (polarized) radio sources. This image-based approach has been used in the past to find pulsars with some success \citep[e.g.][]{nbf+95,bfb+16,fmji16}. In this paper we carry out a pilot study, using the
recently published GMRT 150 MHz all-sky survey \citep[TGSS ADR;][]{ijmf16} and existing images at higher frequencies to search for compact, steep spectrum radio sources within the central 2.5$^\circ$ radius of the gamma-ray excess. In \S\ref{sec:method} we describe our candidate selection method, \S\ref{sec:obs} summarizes our observations of two candidates, data processing, and results, and \S\ref{sec:discuss} presents conclusions.

\section{Method}\label{sec:method}

We generated our initial candidate list from a recently published continuum all-sky survey carried out on the Giant Metrewave Radio Telescope (GMRT) at a frequency of 150 MHz \citep[TGSS ADR;][]{ijmf16}. Pulsars stand out at 150 MHz. With their steep, power-law spectra the true distribution spectral 
indices has a mean $\alpha=-1.4\pm{1}$ \citep{blv13}, but allowing for the biases that affect low frequency surveys, the measured distributions are closer to $\alpha=-1.8\pm0.2$ \citep[where S$_\nu\propto\nu^\alpha$;][]{mkk00,kvh+16,fjmi16}. Thus there is a large frequency leverage arm for spectral index measurements when compared against existing centimeter surveys; those pulsars that may be weak or undetectable at 1.4 GHz are 50-1000$\times$ brighter at 150 MHz. Moreover, steep spectrum radio sources are rare. In \citet{ki08}, fewer than 0.4\% of the radio sources have $\alpha< -1.8$, or less than 1 source per 140 deg$^2$. The only other known discrete radio class with similar spectral slopes are the luminous high redshift galaxies, interesting in their own right but readily distinguished from pulsars from their kpc-size extended structure resolved at arcsecond resolution \citep{md08}.

For this pilot study we concentrated our search on the central 2.5$^\circ$ radius around Sgr A* where the GeV excess is strongest and where there are abundant ancillary data. We found a total of 220 sources in the TGSS ADR catalog at 150 MHz above a threshold of 5$\sigma$. The brightest (faintest) source has a total flux of 35 Jy (43 mJy) and the median value is 250 mJy. As we are interested only in point sources and not extended sources (HII regions, supernova remnants, extragalactic sources, etc.), it would be standard practice to apply a cutoff based on the ratio of the total flux (S$_t$) to peak flux density (S$_p$) following \citet{ijmf16}. However, this approach would likely eliminate real point sources due to the known enhanced scattering over large regions toward the galactic center \citep[e.g.][]{ps14}. For example, a point source whose scattering diameter is \asec{1} at 1 GHz would scale to be \asec{44} at 150 MHz; larger than the \asec{25} restoring beam of the TGSS ADR. We therefore used a more relaxed criterion empirically determined from the data as S$_t$/S$_p\leq{1.51}$. Our final list consists of 166 point-like sources at 150 MHz.


As a initial pass at deriving spectral indices, we compared our source sample with the NRAO VLA Sky Survey (NVSS) catalog at 1.4 GHz \citep{ccg+98} using the \textit{TOPCAT} software package \citep{tay05}. A total of 131 TGSS ADR sources had NVSS counterparts. For those remaining TGSS ADR sources without a NVSS counterpart we visually inspected the NVSS images in order to define an 3$\sigma$ upper flux density limit based on the local noise properties. We reduced this initial candidate list further still by requiring that the initial two-point spectral index $\alpha<-1.4$. This condition was satisfied for 14 spectral index values and 5 spectral index limits, for a total of 19 sources. 

Further refinements to the spectral index measurements of these 19 steep spectrum candidates were made by using \textit{SIMBAD} to search for other imaging surveys of this region. Our primary catalogs were drawn from previous galactic surveys of compact radio sources and include \citet{nlk+04} made at 330 MHz with resolution, $\theta$=\asec{12}$\times$\asec{7} and rms noise, $\sigma$=\mjybeam{1.6}, \citet{zhb+90} at 1.5 GHz ($\theta$=asec{5}, $\sigma$=\mjybeam{1-2}), and \citet{lc98b} at 1.5 GHz ($\theta$=\asec{5}, $\sigma$=\mjybeam{0.4}) and 4.9 GHz ($\theta$=\asec{1.5}, $\sigma$=\mjybeam{0.4}). In addition to our 19 TGSS ADR candidates, we carried out updated \textit{SIMBAD} searches for additional flux density measurements on the five steepest spectrum sources of the 30 pulsar candidates identified in Table 6 of \citet{nlk+04}. New multi-frequency spectral index measurements were derived from these added measurements. Many of the candidates had multiple flux density measurements at 1.4 GHz and flux density variations from one survey to the next were used to identify and eliminate likely resolved sources (although intrinsic variability could not be ruled out in some cases).  Eight compact candidates remain with $\alpha<-1.4$. We chose the steepest spectrum sources with $\alpha<-1.7$ for follow-up. Only two radio sources within a  2.5$^\circ$ radius of Sgr A* satisfied this criterion: TGSS\, J174619.2$-$304010 and TGSS\,J175112.8$-$273723.

\begin{figure}
\includegraphics[width=\columnwidth]{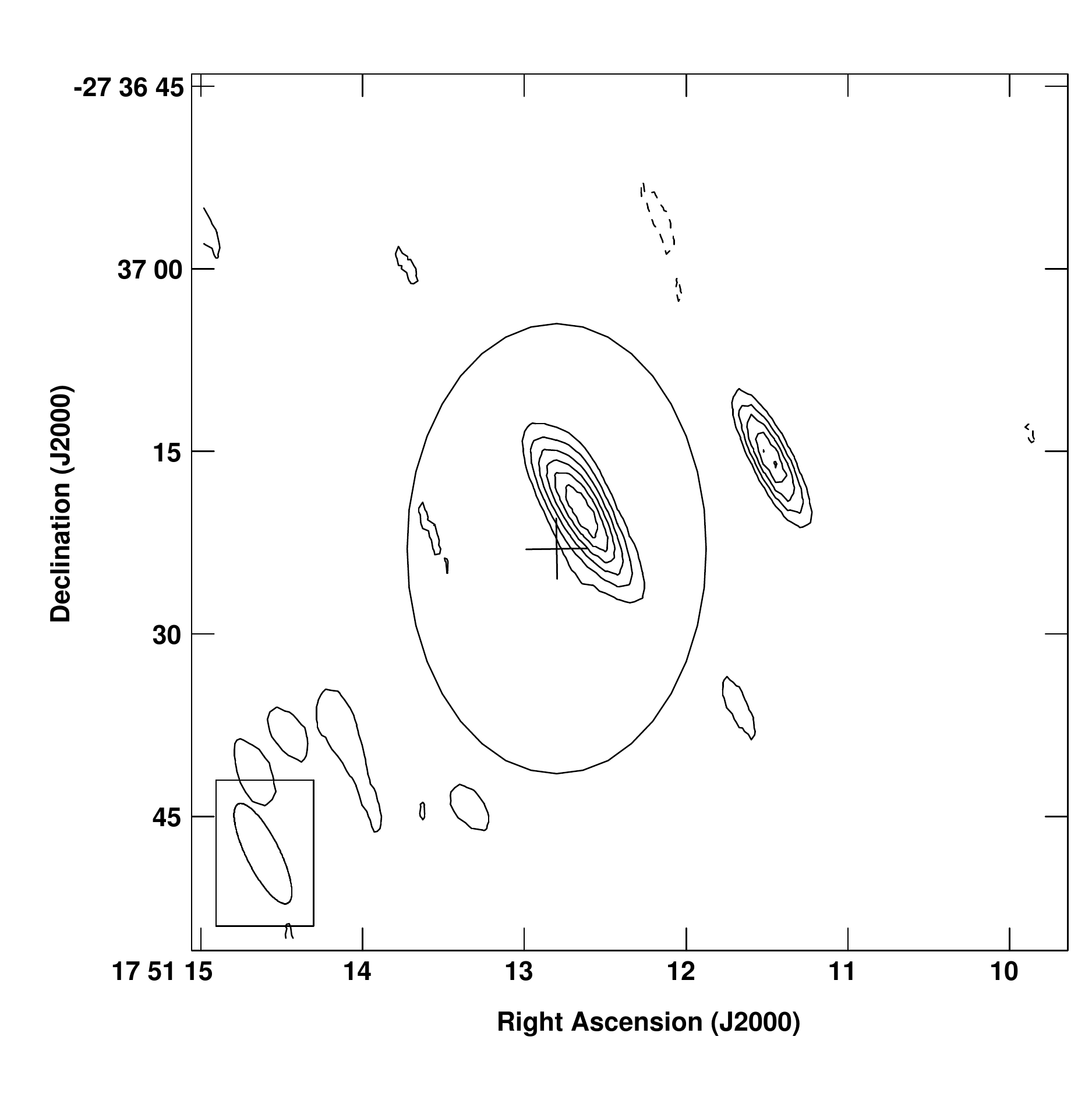}
\caption{A continuum VLA image at 1.5 GHz of two radio sources. The steep spectrum pulsar candidate lies near the center with an ellipse showing the original TGSS ADR beam and a cross indicating the estimate 1$\sigma$ error in this position. A second background radio source lies to the northeast. The VLA synthesized beam size is shown in the lower left corner. The contours are in units of the rms noise of \mujybeam{50} starting at $-$3, 3, 5, 7, 9, 11, 13 and 15.}
\label{fig:psr-agn}
\end{figure}

\section{Observations and Data Analysis}\label{sec:obs}
\subsection{VLA}

The two steep spectrum candidates identified in \S\ref{sec:method} were observed on 2016 July 8 with the Karl G. Jansky Very Large Array (VLA). The observations were carried out over a frequency range of 1-2 GHz with a standard setup of sixteen 64 MHz subbands, with 32 2 MHz channels in each subband (Project code=TDEM0009). The flux density and bandpass calibration used 3C\,48 and phase calibration was carried out with the radio source J1751$-$2524. Data were calibrated using the NRAO pipeline and imaged with the CASA package. The VLA was in its B configuration giving an angular resolution in the images  of approximately \asec{5}.

\begin{figure}
\includegraphics[width=\columnwidth]{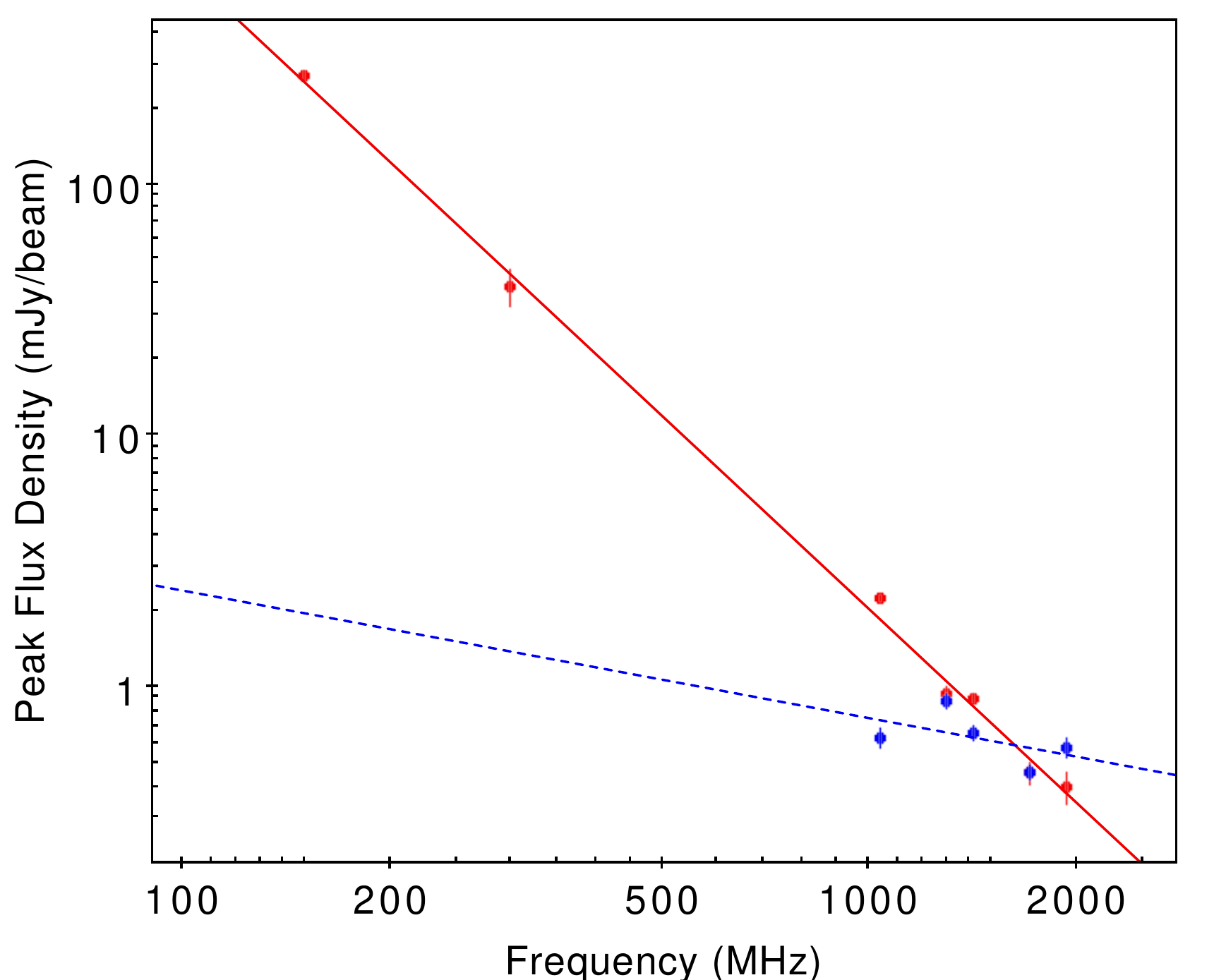}
\caption{Continuum radio spectra of the two VLA point sources, with the steep spectrum pulsar candidate in red and the fainter, flat spectrum source in blue. A least squares fit to the data gives power-law slopes of $\alpha=-0.52\pm{35}$ (blue dashed) and $\alpha=-2.55\pm{0.08}$ (red solid), respectively.}
\label{fig:spectrum}
\end{figure}

At the higher angular resolution of the VLA our first candidate, TGSS\,J174619.2$-$304010, is resolved showing a head-tail morphology typical of extragalactic radio sources. Since we are searching for compact emission from pulsars we will not discuss this source any further. The VLA image for our second candidate, TGSS\,J175112.8$-$273723, is shown in Fig.~\ref{fig:psr-agn}. Two unresolved sources can be seen in this image but only one source is within the original beam of the GMRT detection at 150 MHz. A Gaussian fit to this source yields an improved J2000 position of R.A.=\hms{17}{51}{12.65} and Dec.=\dms{$-$27}{37}{19.8} with an uncertainty of $\pm$\asec{0.3}. We averaged the visibility data into five separate subbands and imaged each, measuring the peak and integrated flux density for both sources. The resulting spectra are also shown in Fig.~\ref{fig:spectrum}. A power-law fit to the peak flux values gives $\alpha=-0.52\pm 0.35$ and $\alpha=-2.55\pm{0.08}$ for the background source and the pulsar candidate, respectively. Fitting to the total flux density instead of the peak flux gives similar slopes. The TGSS source has a spectral slope that lies on the extreme tail of the pulsar spectral index distribution \citep{mkk00,blv13}, while the slope of the other source is more typical of extragalactic sources. 

\begin{table}
\caption{Measured flux values for TGSS\,J175112.8$-$273723. Columns are center observing frequency, peak flux, and total flux density.\label{tab_flux}}
\begin{tabular}{rcc}
\hline			
Freq. & S$_p$ & S$_t$ \\
(MHz) & (mJy/beam) & (mJy) \\
\hline			
150 & 268.4$\pm$30.8 & 260.8$\pm$37.6 \\
330 & 38.8$\pm$6.5 & 48.8$\pm$9.8 \\
1040 & 2.218$\pm$0.066 & 2.170$\pm$0.150 \\
1296 & 0.928$\pm$0.064 & 0.871$\pm$0.147 \\
1424 & 0.889$\pm$0.045 & 0.801$\pm$0.101 \\
1712 & 0.453$\pm$0.044 & 0.625$\pm$0.107 \\
1936 & 0.393$\pm$0.054 & 0.379$\pm$0.151 \\
\hline  
\end{tabular}
\end{table}

\subsection{GBT}\label{obs:gbt}

We observed TGSS\,J175112.8$-$273723 with the GBT (Project code 16B-384) at three epochs and three frequencies, 1.5 GHz (L-band), 2.0 GHz (S-band), and 5.0~GHz (C-band). All observations used the GUPPI backend, and Table~\ref{tab_obs} lists the details of each. While the GUPPI bandwidth was 800~MHz for all observations in the table, the 2.0~GHz GBT receiver has permanent filters which reduce its effective bandwidth to 610~MHz. 

We searched the 2016 October 4 S-band observation for pulsed signals using the Presto software\footnote{\tt http://www.cv.nrao.edu/\~{}sransom/presto/} and a list of 4770 trial dispersion measures (DMs) in the range $0 - 2500$~pc~cm$^{-3}$. A pulsed signal with a period of 2.23~ms was detected at trial $DM = 260.58$~pc~cm$^{-3}$. We identified it as astrophysical based on its wide-band signature and the fact that it exhibits a peak in signal-to-noise vs. trial DM characteristic of dispersed pulsars (see e.g. Figures 3 \& 4 in \citealt{Cordes03}). The period and DM were verified in subsequent S-band and L-band observations; our C-band observation did not yield a detection. The new pulsar is angularly close to the Galactic center with galactic coordinates $l,b$=1.76,$-$0.38, but based on its DM and the NE2001 model of ionized gas in the Galaxy \citep{NE2001-1}, it is a foreground object at a distance of $\sim 4$~kpc. The newer YMW16 model \citep{YMW16} gives a distance of $\sim 3.4$~kpc. Table~\ref{tab_psr} lists the pulsar parameters. 

\begin{figure}
\includegraphics[width=3.5in]{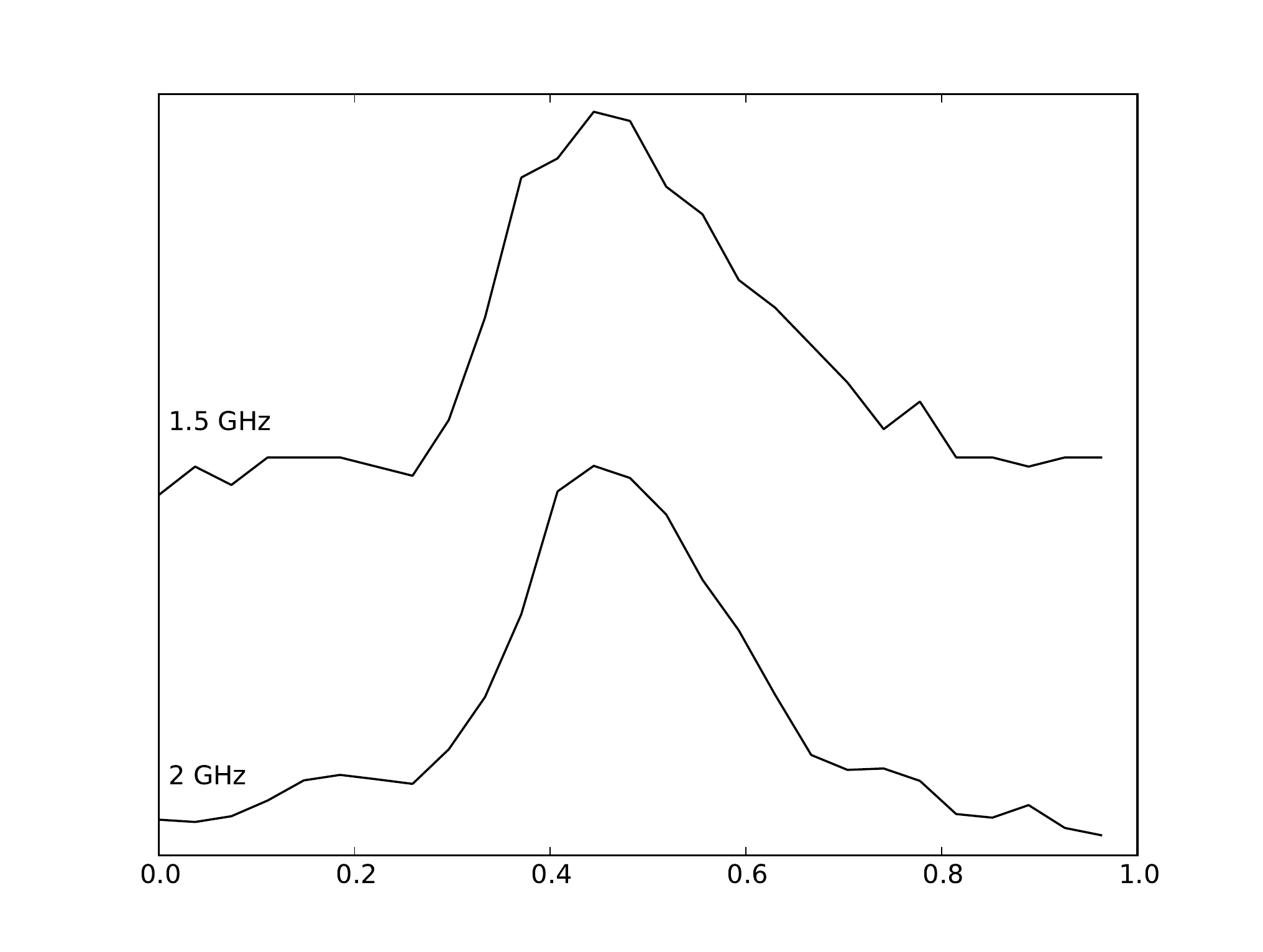}
\caption{Top: normalized average pulse profile of PSR~J1751$-$2737 at 1.5~GHz from our single L-band GBT observation. Bottom: pulse profile at 2.0~GHz obtained by averaging the normalized pulse profiles of our three S-band GBT observations. \label{fig_profs}}
\end{figure}

We processed our GBT observations with {\tt rfifind} and {\tt prepfold} from Presto to excise radio frequency interference (RFI) and produce average pulsar profiles. Figure~\ref{fig_profs} shows the normalized pulse profile (i.e. scaled to a peak of unity) from the single L-band observation and the result from averaging the normalized profiles of the three S-band observations. The FWHM pulse widths are 0.56~ms and 0.52~ms at 1.5~GHz and 2.0~GHz, respectively. 

\begin{table}
\caption{GBT observations of PSR~J1751$-$2737. Columns are center observing frequency, effective bandwidth (the smaller of the backend bandwidth and unfiltered receiver bandwidth), sampling time, gain, system temperature, integration time, and the presence of a pulsed signal detection. Tsys includes instrumental contributions as well as the cosmic microwave background and atmospheric emission.\label{tab_obs}}
\begin{tabular}{lccccccc}
\hline			
Date & $f_{\rm c}$ & $\Delta\nu$ & $\Delta t$ & G & $T_{\rm sys}$ & $T_{\rm obs}$ & Detect? \\
(2016) & (GHz) & (MHz) & ($\mu$s) & (K/Jy) & (K) & (hrs) & \omit \\
\hline			
Oct.\,4 & 2.0 & 610 &  81.92 & 2.0 & 20 & 0.79 & Y\\
Oct.\,11 & 2.0 & 610 &  81.92 & 2.0 & 20 & 1.09 & Y\\
Oct.\,11 & 5.0 & 800 &  81.92 & 1.9 & 18 & 2.89 & N \\
Oct.\,23 & 1.5 & 800 &  81.92 & 1.9 & 20 & 0.96 & Y\\
Oct.\,23 & 2.0 & 610 &  81.92 & 2.0 & 20 & 1.00 & Y\\
\hline  
\end{tabular}
\end{table}

\begin{table}
\caption{Properties of PSR~J1751$-$2737: Coordinates are from VLA imaging data, all other values are derived from GBT time domain data.\label{tab_psr}}
\begin{tabular}{ll}
\hline			
Parameter & Value\\
\hline			
Name & J1751$-$2737 \\
Right Ascension (J2000) & 17${\rm ^h}$ 51${\rm ^m}$ 12${\rm ^s}$.65(2)\\
Declination (J2000) & $-$27\degrees 37'19.8(3)"\\
Rotation period $P$(ms) & 2.23 \\
Dispersion measure DM(pc~cm$^{-3}$) & 260 \\
NE2001 distance $D$ (kpc) & 4.0 \\
YMW16 distance $D$ (kpc) & 3.4 \\
$S_{\rm 1.5GHz}$ (mJy) & $>0.32$ \\
$S_{\rm 2GHz}$ (mJy) & 0.30 \\
$S_{\rm 5GHz}$ (mJy) & $<0.04$ \\
Spectral index $\alpha$ ($S \propto \nu^\alpha$) & $< -2.2$ \\
$\sigma_{\rm 1.5GHz}$ (ms)  & 0.15 \\
$\tau_{\rm s, 1.5GHz}$ (ms) & 0.30 \\
$\sigma_{\rm 2GHz}$ (ms)  & 0.18 \\
$\tau_{\rm s, 2GHz}$ (ms) & 0.19 \\
\hline  
\end{tabular}
\end{table}

We fit the two normalized average profiles with a Gaussian convolved with an exponential with a decay time corresponding to the scattering broadening time, $\tau_s$. The S-band best fit Gaussian $\sigma = 0.18$~ms and $\tau_s = 0.19$~ms, and the L-band best fit $\sigma = 0.15$~ms and $\tau_s = 0.30$~ms. For comparison, we also computed the expected $\tau_s$ for the two observing frequencies and the pulsar DM using the NE2001 model of ionized gas in the Galaxy\footnote{\tt https://www.nrl.navy.mil/rsd/RORF/ne2001}. This yielded $\tau_s$ values of 0.07 and 0.26~ms for S-band and L-band, respectively. Our results are within the uncertainties of models estimating $\tau_s$ based on measured pulsar DM (\citealt{NE2001-2}, \citealt{Bhat04}).

We estimate the flux density of PSR~J1751$-$2737 based on the radiometer equation, the average pulse profile of each detection, the parameters in Table~\ref{tab_obs}, and the background sky temperature $T_{\rm sky}$ at each frequency. In order to find $T_{\rm sky}$ at each of our observing frequencies, we scale from the observed value of 3.1~K at 2695~MHz \citep{Reich90} to each frequency using a power law with an index of $-2.6$ \citep{RR88}.\footnote{PSR~J1751$-$2737 is outside the \cite{RR88} spectral index map by $\sim 7\degrees$; however, the spectral index of $T_{\rm sky}$ does not vary significantly on that spatial scale in the vicinity of the pulsar.} We obtain $T_{\rm sky,1.5GHz} = 14$~K, $T_{\rm sky,2GHz} = 7$~K, and $T_{\rm sky,5GHz} = 0.6$~K. 

From our 1.5~GHz observation we obtain a period-averaged flux density of 0.32~mJy. However, the observation was plagued by strong and pervasive RFI that was difficult to excise without masking most of the data. This flux density value should be taken as a lower limit. From our 2.0~GHz observations we obtain period-averaged flux densities of 0.35, 0.16, and 0.38~mJy. While the S-band receiver bandwidth is significantly cleaner than the L-band bandwidth, the outlying middle measurement may be due to imperfectly cleaned RFI. Regular observations during our planned timing campaign of the new MSP will be able to distinguish between this possibility and propagation effects such as refractive scintillation. The average flux density of the three 2.0~GHz observations, 0.30~mJy, agrees very well with the extrapolation of the VLA measurements in Table \ref{tab_flux} and Fig.~\ref{fig:spectrum}. From our non-detection at 5.0~GHz, we calculate an upper limit of 0.04~mJy, assuming a detection threshold of 10$\sigma$. From the C-band upper limit and the average S-band flux density, we calculate an upper limit on the spectral index of the pulsar, $\alpha < -2.2$, again in good agreement value derived from interferometric data. 

\section{Discussion and Conclusions}\label{sec:discuss}

PSR\,J1751$-$2737 appears to be an isolated recycled pulsar located within the disk of our Galaxy. A binary pulsar would have been expected to show some change in its period over the three week interval that it was observed with the GBT while none was observed (Table \ref{tab_psr}). However, until longer-term timing is conducted, a long period binary or one with a low mass fraction cannot be ruled out. The pulsar's radio luminosity at 1.4 GHz, defined in the usual way as L$_{1.4}$=S$_t\times$d$^2$ and using the NE2001 distance is 13 mJy kpc$^2$. This value of L$_{1.4}$ is on the high end of the lognormal luminosity function for MSPs in globular clusters \citep{blc11} and is a bright outlier in the existing sample of isolated MSPs in the disk \citep{bbb+13}. The dispersion measure distance and the limits on temporal scattering (\S\ref{obs:gbt}) favor an origin in the galactic plane. For the pulsar to be located in the bulge at $\sim$8.5 kpc the NE2001 model predicts a DM=655 pc cm$^{-3}$ and $\tau_s$ at 2 GHz of 2.4 ms, values much higher than we observe. Thus PSR\,J1751$-$2737 is not a member of the putative bulge population that is thought to be responsible for the \fermi excess. There are no discrete gamma-ray sources in this direction either in the 3FGL catalog  or in catalogs of the galactic center region \citep{aaa+15,aaa+16}. A search for gamma-ray pulsations may still be worthwhile since the complex background models in this direction make it difficult to robustly identify discrete sources.  

Most past sensitive centimetre radio pulsar surveys did not search a region large enough to detect PSR\,J1751$-$2737. Owing to the large amount of telescope time required for high frequency single-dish surveys, some searches have concentrated within a one degree radius centered on SgrA* \citep{jkl+06,dcl09}, while others have focused deep integrations within the central few parsecs around SgrA* \citep{mkfr10,sbb+13,ekk+13}. The lone exception is the on-going High Time Resolution Universe (HTRU) survey of the southern sky with the Parkes multi-beam receiver at 1.4 GHz, which has covered the significant parts of the \fermi excess area. In particular, the low latitude component of the southern HTRU survey is making hour-long integrations with the goal of searching the galactic plane in latitude $\vert{b}\vert<{3.5}^\circ$ and longitude 30$^\circ<{l}<280^\circ$ \citep{kjs+10}. HTRU has searched the direction toward PSR\,J1751$-$2737 but the pulsar does not appear in recent lists of detections \citep{ncb+15}. This is somewhat surprising given the experimental parameters of the survey and the similarity between the GBT and Parkes sensitivities \cite[see Table \ref{tab_obs} and][]{kjs+10}. A non-detection above threshold could be explained if PSR\,J1751$-$2737 was beyond the half-power point of a single feed, especially if it was located near an outer feed where the telescope gain is lower. In any case, with the position and period from Table \ref{tab_psr} a renewed HTRU archival search would be worthwhile. When compared to the GBT data, earlier HTRU measurements could be used to help constrain \.{E} and other parameters.

While PSR\,J1751$-$2737 is not a bulge MSP, this pilot project has shown that interferometric imaging observations have a role in constraining the contribution of young and recycled pulsars to the gamma-ray excess around the galactic center. \citet{cdd+16} have looked at the prospects for detecting a bulge population of MSPs in some detail. They show that while existing deep surveys such as the HTRU are not well-suited to bulge detections, future large area radio pulsation searches have the potential to yield dozens of detections. The disadvantage of this direct approach, however, is that it is currently prohibitively expensive, either in telescope time or in computational  resources. Here we used a hybrid approach. We undertook interferometric imaging at low frequencies where a single 15-minute integration at 150 MHz covered a field of view of five square degrees. Pulsar candidates were identified as steep spectrum point sources. This approach is sensitive to only phase-averaged flux density. One limitation of this method is that is it is not sensitive to flat spectrum pulsars. In the current pulsar catalog one third of all pulsars have a spectral index $\alpha>-1.6$. This disadvantage is balanced against other advantages. The method makes no assumptions about the pulsar period, the dispersion measure, temporal scattering or binarity.  These quantities were searched over from a time series taken in a deep single-pointing pulsation search carried out at higher frequencies.

We can make some rough comparisons between a direct pulsation search and the hybrid technique. \citet{cdd+16} describe a pulsation search with a 100-m class single dish radio telescope at 1.4 GHz. Their region with the highest yield is five degrees above (or below) the Galactic center. The surface density of radio bright bulge MSPs (defined as $\geq$10 $\mu$Jy at 1.4 GHz) is still large (4.7$\pm$1.5 deg$^{-2}$) but the sky brightness temperature, the scattering and dispersive smearing are all sharply reduced compared to the Galactic center. An a 250-hr experiment they show that a 100-m can carry our a sensitive survey of a 2$^\circ\times{2}^\circ$ area, detecting 1.7 bulge MSPs or 0.43 detections per square degree. A hybrid approach would begin with a deep interferometric image of the same region. For example, a single 1-hr pointing of the upgraded GMRT 325 MHz could achieve a thermal noise\footnote{N. Kanekar, priv. comm.} (5$\sigma$) of 15 $\mu$Jy beam$^{-1}$ over a field of view of 1.4 deg$^{2}$.  Pulsars are ten times brighter at 325 MHz than at 1.4 GHz, so this experiment would detect {\it all} radio bright bulge MSPs. Since approximately 2/3 of these would be selected based on their steep spectrum, imaging over the GMRT field of view result in 4.5 MSP candidates, or 3.1 detections per square degree. [Note that the smaller dishes of the VLA would require more integration time to reach similar noise levels but they would image a field of view three times larger.] Follow-up single-dish pulsation searches would require several hours to confirm pulsations from each of these 4.5 candidates but they would be made at frequencies higher than 1.4 GHz, where the scattering and dispersive effects are reduced by $\lambda^4$ and $\lambda^2$, respectively. These numbers are only approximate but they show that the hybrid technique might be used to get an order of magnitude yield of MSP bulge detections, with a concomitant reduction in total observing time. 

This hybrid approach could be used in the future to efficiently search for a bulge population of pulsars. Imaging searches for compact steep spectrum source toward the galactic center region have been carried out in the past \citep{lc98b,nlk+04}, but a new generation of low temperature, broadband feeds
motivates a re-thinking of new surveys. Instantaneous, wide bandwidth measurements of flux densities, while subject to some uncertainties \citep{rbo16}, are preferable to measuring the spectral index by comparing images from different telescopes with large differences in angular resolution and different calibration schemes, and often taken at different times so that variability can produce false detections. For example, many of the initial steep spectrum candidates identified in \citet{nlk+04} turn out to be false positives due to resolution effects or variability. TGSS\,J175112.8$-$273723 on the other hand could have been identified as a strong pulsar candidate based solely on the follow-up VLA measurements taken from 1 to 2 GHz (Table \ref{tab_flux}). 

The VLA L-band system (1-2 GHz) and the upgraded P-band system at the GMRT (250-500 MHz) have the large fractional bandwidths ($\delta\nu/\nu$=33\%) required to detect steep spectrum candidates in-band. They also have the wide instantaneous field of view needed to efficiently image the entire region of the \fermi excess with arcsecond resolution. Our choice of 150 MHz for the pilot study was sub-optimal since the bandwidth is narrow ($\delta\nu/\nu$=10\%) while the temperature of the diffuse sky background is large and rises as $\nu^{-2.6}$, i.e. steeper than most pulsars. There is also free-free absorption from the patchy, ionized gas at the galactic center that attenuates pulsar emission at this frequency \citep{ps14}. With the proper choice of channel widths and integration times for the interferometer setup, pulsar candidates could be further identified based on their diffractive scintillations as measured in variance images \citep{djb+16}. In the mean time, before a full hybrid survey is carried out, we agree with \citet{lkkd16} and \citet{cdd+16} that a search should be made for bulge pulsar candidates toward the "hotspots" identified by \citet{bkw16} and \citet{lls+16}. The detection of PSR\,J1751$-$2737 shows that deep imaging-based continuum with targeted pulsation searches offers a complementary approach to existing blind pulsation surveys.

\section*{Acknowledgments}
This research has made use of NASA's Astrophysics Data System (ADS) and of the SIMBAD database, operated at CDS, Strasbourg, France. This study has made use of data obtained with the GMRT, run by the National Centre for Radio Astrophysics of the Tata Institute of Fundamental Research. The National Radio Astronomy Observatory is a facility of the National Science Foundation operated under cooperative agreement by Associated Universities, Inc. We thank the directors of the VLA (Mark McKinnon) and the GBT (Karen O'Neil) for providing Director's Discretionary Time to follow up on our candidate sources. DB thanks Drs A. Corsi, T. Maccarone and B. Owen at TTU for their support and advice.

{\it Facilities:} VLA, GBT, Fermi (LAT).

\end{document}